\documentclass[aps,prl,twocolumn, epsfig]{revtex4}
\usepackage[dvips]{graphics}
\usepackage{graphicx}
\usepackage{amsfonts}
\usepackage{amssymb}
\usepackage{amsmath}
\begin{document}
\title{Quasi-crystal structures of
Bose-Einstein condensates}
\author{Mario Salerno}
\affiliation{Dipartimento di Fisica "E.R. Caianiello",
 Universit\'a di Salerno, via S. Allende, I-84081 Baronissi (SA),\\
         Consorzio Interuniversitario per le Scienze Fisiche della
         Materia (CNISM), Unita' di Salerno,
         \\ Istituto Nazionale
         di Fisica Nucleare (INFN), Gruppo Collegato di Salerno, Italy.}
\begin{abstract}
Quasi-crystal structures of Bose-Einstein condensates which
correspond to a Penrose tiling of the plane are discussed.
\end{abstract}
\maketitle

We discuss quasi-crystals structures of Bose-Einstein condensates
(BEC) which correspond to a Penrose tiling of the plane. The
underlying model is the repulsive 2D Gross-Pitaevskii equation
\begin{equation}
i\psi_{t}+\left[ \partial_{xx} +\partial_{yy} +V_{ol}({\bf
r})-|\psi|^{2}\right] \psi=0 \label{gpe}
\end{equation}
with the following optical lattice
\begin{equation}
\label{OL} V_{ol}(x,y)=\varepsilon \sum_{i=1}^n \cos[\pi (s_i x +
c_i y)],
\end{equation}
where $s_i=\sin(\frac{2 \pi}{n} i)$, $c_i=\cos(\frac{2 \pi}{n} i)$
and $\varepsilon$ denotes the strength of the lattice. For
$n=3,4$, this potential originates a Bravais lattice with
three-fold and four-fold symmetry, respectively. For $n=5$ it
produces a lattice with an impossible five-fold quasi symmetry
which correspond to the Penrose tiling of the plane
\cite{penrose}. To show this, we first study the interference
patterns which are created when an initial gaussian distribution
of matter
\begin{equation}
\label{ic} \psi(x,y)=A\exp \left( - a (x^2+y^2)\right)
\end{equation}
is released in such quasi-periodic potential. To allow the matter
to expand in the structure, we consider the case of small
potential amplitude and relatively small number of atoms. The
result are shown in the following figures. In Fig.1 the
interference pattern created after a time $t=0.5$ is depicted. In
Fig. 2 we show the same pattern of Fig. 1 with lines joining the
points where the peak intensity is higher superimposed. The
formation of the Penrose tiling of the plane becomes then evident,
this providing an example of BEC quasi-crystal \cite{santos}.
Taking $n>6$ in Eq. (\ref{OL}) one can generate other types of
planar quasi-crystals (i.e. with heptagonal (n=7), octagonal
(n=8), etc., quasi-symmetries).

Next we investigate the possibility to form  localized states of
soliton type  in 2D quasi-crystals. As a result we show that
soliton states  can indeed be formed. Although this is expected,
due to the the trapping features of the potential (especially for
high values of $\varepsilon$),  a main difference with
gap-solitons of periodic lattices is that in quasi-crystals the
shape and properties of the soliton strongly depends on the
position in the lattice. Due to the lacking of a strict
periodicity in the system one may expect the existence of a large
variety of localized excitations in BEC quasi-crystals.

\begin{figure}[th]
\centerline{\includegraphics[width=4cm,height=4cm,clip]{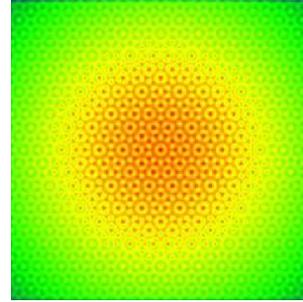}}
\vspace*{0.5cm} \caption{BEC interference pattern at time $t=0.5$
obtained from direct simulations of Eq. (\ref{gpe}) with the
potential in Eq. (\ref{OL}) for $n=5$ and $\protect\varepsilon
=0.2$. The initial condition was taken as in Eq. (\ref {ic}) with
$a=2.2*10^{-4}$ and $A=0.264$. The length of the system is
$L=28\pi$} \label{fig1}
\end{figure}

\begin{figure}[th]
\centerline{\includegraphics[width=4cm,height=4cm,clip]{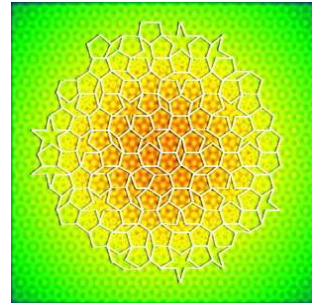}}
\vspace*{0.5cm} \caption{Same as in Fig.1 but with lines  joining
points where the matter wave intensity is higher (big red spots in
Fig.1 ), superimposed. The tiling lines were drawn by hand using
an art-work program. } \label{fig2}
\end{figure}


\begin{references}
\bibitem{penrose}see R. Penrose, The emperor's new mind, Vintage,
1990.
\bibitem{santos}BEC in optical quasi-crystals has been  recently
reported also by L. Sanchez-Palencia and L. Santos (see
cond-mat/0502529)
\end{references}
\end{document}